\documentclass{amsart}
\usepackage{graphicx}
\usepackage{epsf}
\usepackage[italian]{babel}
\def\be{\begin{equation}}
\def\ee{\end{equation}}
\def\ba{\begin{array}}
\def\ea{\end{array}}
\def\bea{\begin{eqnarray}}
\def\eea{\end{eqnarray}}
\def\drm{{\mathrm d}}

\def\rot{{\mathrm{rot}}\,}
\def\div{{\mathrm{div}}\,}

\begin{document}

\vspace{-4truecm} %
{}\hfill{DSF$-$20/2006} %
%%physics/0607099 %
\vspace{1truecm}

\title{Un manoscritto inedito in {\it francese} di Ettore Majorana}%
\author{S. Esposito}%
\address{{\it S. Esposito}: Dipartimento di Scienze Fisiche,
Universit\`a di Napoli ``Federico II'' \& I.N.F.N. Sezione di
Napoli, Complesso Universitario di M. S. Angelo, Via Cinthia,
80126 Napoli ({\rm Salvatore.Esposito@na.infn.it})}%

%\thanks{}%
%\subjclass{}%
%\keywords{}%

%\date{}%
%\dedicatory{}%
%\commby{}%
%----------------------------------------------------------------

\begin{abstract}
Nel presente lavoro viene data notizia del ritrovamento delle
ultime carte mancanti del {\it fascicolo Senatore}, consegnato da
Majorana ad una sua studentessa a Napoli, poco prima di
scomparire, nel 1938. Il manoscritto in questione, conservato alla
Domus Galilaeana a Pisa, fu redatto in lingua {\it francese},
probabilmente in preparazione per una conferenza a Leningrado (o a
Kharkov) del 1933 (o 1934), alla quale inizialmente Majorana fu
invitato, ma a cui, per\`o, non partecip\`o. Il testo ritrovato,
che tratta di elettrodinamica quantistica con l'uso del formalismo
della quantizzazione dei campi, viene qui riportato in traduzione
per la prima volta, fornendone un accurato inquadramento sia
storico che scientifico.
\end{abstract}

\maketitle

%----------------------------------------------------------------

\section{Le ultime carte di Majorana}

\noindent Ettore Majorana, nato il 5 agosto 1906, insegn\`o dal
gennaio al marzo del 1938 all'Universit\`a di Napoli, avendo
ottenuto ``fuori concorso'' la cattedra di fisica teorica presso
l'Istituto di Fisica di tale ateneo \cite{Recami}. L'interesse per
il corso svolto dal grande fisico catanese, l'unico effettivamente
tenuto davanti a degli studenti \cite{DeGregorio}, \`e gi\`a stato
pi\`u volte evidenziato (si veda, per esempio, \cite{Moreno} e
\cite{Weyl}), e risiede sostanzialmente nel carattere
d'avanguardia dato da Majorana al suo insegnamento della meccanica
quantistica. La vicenda storica e personale del personaggio,
invece, ruota attorno alla sua misteriosa scomparsa, avvenuta alla
fine di marzo del 1938 proprio durante il suo soggiorno a Napoli.
Il venerd\`i 25 marzo, il giorno dopo aver svolto la sua
ventunesima lezione, secondo quanto racconta una sua studentessa,
Gilda Senatore, $\ll$Majorana contrariamente a quanto di solito
faceva [quando non era prevista alcuna lezione] venne in Istituto
e vi si trattenne soltanto pochi minuti... Dal corridoio che
immetteva nell'auletta in cui mi trattenevo scrivendo, mi chiam\`o
per nome: `Signorina Senatore...'; non entr\`o ma si trattenne nel
corridoio; lo raggiunsi ed egli mi consegn\`o una cartella chiusa
dicendomi: `ecco, prenda queste carte, questi appunti... poi ne
riparleremo'; poi and\`o via e voltandosi ripet\`e: `poi ne
riparleremo'$\gg$ \cite{Senatore}. Quella cartella conteneva gli
appunti del corso di fisica teorica preparati per i suoi studenti
da Majorana che, in quella occasione, prima di scomparire, volle
definitivamente consegnare ad una sua studentessa. Ci\`o che
rimane degli originali di quegli appunti fu pubblicato in stampa
anastatica nel 1987 \cite{bibliopolis87}. Quando, tuttavia, la
Senatore vision\`o tale pubblicazione ossev\`o che $\ll$mancano
alcuni capitoli delle lezioni, il cui testo mi fu consegnato,
completo persino degli appunti di quella che il professore avrebbe
dovuto tenere il giorno successivo alla sua scomparsa. Manca
ancora un esiguo gruppo di fogli, scritti anche essi in originale
ed in maniera ordinata come gli altri, ma non facenti parte delle
lezioni gi\`a tenute$\gg$ \cite{Senatore}. Il percorso degli
originali di Majorana dal 1938 fino al 1966, quando Edoardo Amaldi
li deposit\`o presso la Domus Galilaeana a Pisa insieme con le
altre carte di Majorana, si \`e sviluppato in modo tortuoso, come
evidenziato altrove \cite{Moreno}, per cui potrebbe non arrecare
meraviglia il fatto che alcuni ``capitoli'' siano andati
``persi''. Oggi, tuttavia, si ritiene di aver finalmente
ricostruito l'intero {\it corpus} degli appunti del corso di
lezioni di fisica teorica \cite{bibliopolis06}, grazie al recente
ritrovamento del Documento Moreno \cite{Moreno}, con il quale sono
emersi gli appunti delle sei lezioni del detto corso di cui non si
disponeva del manoscritto originale.

Riguardo, invece, agli appunti di quella che la Senatore ha
interpretato come la lezione che ``il professore avrebbe dovuto
tenere il giorno successivo alla sua scomparsa'', un'analisi molto
recente \cite{path} ha portato a concludere che essi non si
riferiscono affatto ad una possibile lezione del corso di fisica
teorica, ma piuttosto al testo di una conferenza o seminario
generale, forse tenuto nello stesso Istituto di Fisica partenopeo,
ma certamente non rivolto a studenti bens\`i a ricercatori.

\section{Il fascicolo Senatore}

\noindent Sul rimanente ``esiguo gruppo di fogli'' non si \`e
stranamente svolta, finora, alcuna indagine fruttuosa, sebbene sia
passato un lungo periodo temporale sia dal deposito delle carte di
Majorana a Pisa (1966) che, soprattutto, dalla pubblicazione della
stampa anastatica degli originali degli appunti del corso di
Majorana (1938). Tale indagine \`e stata condotta da chi scrive
principalmente sulla base delle fonti d'archivio della Domus
Galilaeana, e il risultato viene qui di seguito presentato per la
prima volta.

Il fascicolo del corso tenuto a Napoli, consegnato alla Senatore
il 25 marzo 1938 e conservato ora a Pisa contiene i seguenti
documenti.

Innanzitutto \`e presente l'originale della lettera di Gilberto
Bernardini ad Amaldi del 2 dicembre 1964 (riprodotta in
\cite{bibliopolis87}; in essa Bernardini risponde ad Amaldi circa
la sua insistente richiesta di notizia sugli appunti del corso di
Majorana) che, evidentemente, \`e stata aggiunta successivamente
al fascicolo da Amaldi.

Il secondo documento\footnote{La divisione in ``documenti'' \`e
fatta qui per comodit\`a di linguaggio, ma non corrisponde ad
alcuna reale divisione. Infatti nel fascicolo in esame sono
presenti solo fogli sparsi, secondo quanto descritto qui nel
testo.} presente nel fascicolo \`e composto da 22 fogli (di 4
pagine ciascuno) approssimativamente di formato A4 e a struttura
rigata di tipo ``commerciale'' (ossia la struttura \`e di tipo
rettangolare). Essi contengono il testo degli appunti originali di
10 lezioni del corso \cite{bibliopolis06}.

Il terzo documento \`e invece composto da 3 fogli (di 4 pagine
ciascuno) approssimativamente dello stesso formato dei precedenti,
ma con una struttura ``a quadretti''. Il testo in esso presente
\`e scritto con una penna (ad inchiostro nero, come per il
precedente documento) diversa da quella utilizzata per i fogli
precedenti, essendo il suo tratto pi\`u fine. Il contenuto si
riferisce a quanto menzionato sopra a proposito della conferenza o
seminario generale preparato da Majorana; per ulteriori
approfondimenti si rimanda a \cite{path}. Per possibili futuri
riferimenti, tale documento composto da 3 fogli verr\`a indicato
come ``{\it excerpta} Senatore I''.

Inaspettatamente, il fascicolo della Domus Galilaeana contiene
anche un documento, composto da 2 fogli (il primo di 4 pagine, il
secondo di sole 2 pagine), anch'essi approssimativamente di
formato A4 e con struttura ``a quadretti''. Tale documento,
denominato qui ``{\it excerpta} Senatore II'', \`e tuttavia
completamente differente dagli {\it excerpta} Senatore I (e dagli
appunti delle lezioni) sia per il testo, che \`e scritto in {\it
francese} e non in italiano, che per la penna adoperata, che
utilizza un inchiostro (nero) pi\`u scuro e il tratto \`e pi\`u
largo.

Il contenuto del testo di tale manoscritto, riportato in
traduzione in lingua italiana in appendice, riguarda
un'applicazione alla teoria delle lacune di Dirac del formalismo
della seconda quantizzazione ({\it excerpta} Senatore IIa) e
alcuni argomenti di elettrodinamica quantistica ({\it excerpta}
Senatore IIb e IIc), che verranno inquadrati teoricamente nel
successivo paragrafo. \`E per\`o interessante notare subito che,
nonostante l'apparente diversit\`a di contenuti, l'intero testo
francese presente nei 2 fogli sembra essere stato stilato in
un'unica occasione.

\section{Il contesto teorico di riferimento}

\noindent Nel quadro generale della meccanica ondulatoria,
introdotta sostanzialmente da Erwin Schr\"odinger, l'esistenza di
un `quanto di materia'', come l'elettrone, poneva gli stessi
problemi teorici di quelli dell'esistenza dei quanti di luce nel
contesto della teoria elettromagnetica di Maxwell, come rilevato
esplicitamente da Werner Heisenberg nel 1929 \cite{Hei29}. Se,
infatti, il problema fondamentale associato all'esistenza dei
fotoni era quello di spiegare il fenomeno dell'interferenza, per
gli elettroni era quello di ricavare la quantizzazione della
carica elettrica a partire dalla funzione d'onda di Schr\"odinger.

Come noto, la soluzione di tali difficolt\`a venne con
l'introduzione del procedimento di quantizzazione dei campi (o
``seconda quantizzazione'') che, dopo un lavoro preliminare di
Pascual Jordan del 1926 \cite{Jordan26}, fu applicato per la prima
volta da Paul A. M. Dirac nel 1927 \cite{Dirac27} alla
quantizzazione del campo elettromagnetico. Il passo successivo fu
poi quello di considerare particelle con massa diversa da zero, e
ci\`o fu realizzato poco dopo ad opera di Jordan e Oskar Klein
\cite{JK27}, che generalizzarono il metodo di quantizzazione di
Dirac a campi bosonici. Il procedimento da adottare nel caso di
particelle fermioniche, che doveva assicurare l'accordo con il
principio di esclusione di Pauli, richiese invece ``l'invenzione''
degli anticommutatori, come evidenziato da Jordan e Eugene Wigner
nel 1928 \cite{JW28}.

Il punto cruciale della questione fu, evidentemente, la
realizzazione che i campi d'onda (di Schr\"odinger, di Dirac o di
Maxwell) non rappresentano singole particelle ma, al contrario, un
numero arbitrario di particelle, ed \`e a questo riguardo
particolarmente significativo il titolo dato da Jordan e Klein al
loro fondamentale articolo, {\it Sul problema a molti corpi della
teoria dei quanti}.

L'inclusione dell'interazione tra luce e materia nei procedimenti
di seconda quantizzazione fu considerato organicamente nel
1929-1930 da Heisenberg e Wolfgang Pauli \cite{HP29}, che
elaborarono una prima teoria dell'elettrodinamica quantistica
facendo uso del formalismo lagrangiano. In tale teoria, tuttavia,
cominciarono ad affermarsi prepotentemente alcuni problemi
fondamentali, quali l'insorgenza di una energia coulombiana di
autointerazione per l'elettrone infinita e il problema degli stati
ad energia negativa (ossia la transizione tra stati con $E=+ m
c^2$ a stati con $E=- m c^2$) che, sebbene impegnarono le forze
dei migliori fisici teorici dell'epoca, rimasero irrisolti per un
certo periodo. In particolare il problema degli stati ad energia
negativa si acu\`i quando Ivar Waller \cite{Waller30} dimostr\`o
la necessit\`a della loro esistenza per ottenere il limite
classico di Thomson nella diffusione dei raggi X da parte degli
atomi. La teoria di Dirac delle ``lacune'' \cite{Dirac30}, prima
con l'interpretazione di queste in termini di protoni (seguendo
una congettura di Hermann Weyl \cite{Weyl29}) e poi con la
predizione e scoperta dei positroni, offr\`i come ben noto una
semplice e interessante soluzione al problema, introducendo il
concetto di ``vuoto'' (o, usando le parole di Dirac, ``stato
normale di elettrificazione'') come un mare di elettroni nel quale
tutti gli stati ad energia positiva sono vuoti, mentre tutti
quelli ad energia negativa sono occupati.

L'accettazione della teoria delle lacune, tuttavia, fu inizialmente
molto variegata all'interno della comunit\`a dei fisici, se ancora
nel 1933, dopo la scoperta dell'esistenza dell'``antielettrone'',
Pauli espresse in molte occasioni le sue perplessit\`a su di essa
\cite{Pauli33}. Un punto di vista completamente opposto fu invece
adottato da Heisenberg che, contrariamente a Pauli, vedeva molto
favorevolmente nella teoria di Dirac una sostanziale simmetria tra
processi che coinvolgono elettroni e quelli con positroni
\cite{HS33}, un motivo che verr\`a ripreso formalmente da Majorana
nel suo famoso articolo sulla {\it Teoria simmetrica dell'elettrone
e del positrone} \cite{elpos} (la cui genesi, probabilmente, avvenne
proprio gi\`a nel 1933, quando Majorana si trovava da Heisenberg a
Lipsia). Lo stesso Heisenberg, infatti, nel 1931 elabor\`o
un'applicazione, particolarmente interessante per i nostri scopi, in
cui considerava la simmetria tra lacune ed elettroni in un livello
atomico occupato o in una banda di energia occupata di un cristallo
\cite{Hei31} (ossia: un'applicazione ad un caso ``concreto'', in
contrapposizione al ``vuoto'' di Dirac). In tale lavoro, Heisenberg
ottenne una hamiltoniana di seconda quantizzazione espressa in
termini delle lacune (si veda l'Eq. (15) di Ref. \cite{Hei31}), in
cui si evidenziavano dei termini operatoriali aggiuntivi di
singolaparticella derivanti dalle commutazioni di operatori a due
particelle. L'applicazione al caso di un livello atomico di $N$
elettroni contenente, per\`o, solo $n$ elettroni \`e studiato nella
prima parte dell'articolo, e porta ad una equazione d'onda (Eq. (17)
di Ref. \cite{Hei31}) per $N-n$ lacune, mentre la simmetria
elettrone-lacuna nel caso di un metallo con un effetto Hall
``anomalo'' \`e trattato nella seconda parte di esso. Come gi\`a
accennato, Heisenberg usa il formalismo della seconda
quantizzazione, facendo espresso riferimento sia all'articolo
generale di Jordan e Klein del 1927 \cite{JK27} che a quello di
Jordan e Wigner del 1928 \cite{JW28} in cui vengono appropriatamente
introdotti gli anticommutatori per i fermioni (nel caso specifico,
elettroni e lacune).

Negli {\it excerpta} Senatore IIa Majorana segue sostanzialmente
la traccia di Heisenberg in \cite{Hei31}, ma non \`e affatto
chiaro se l'autore si riferisce ad una applicazione ``concreta''
(come quella di Heisenberg) oppure, molto pi\`u interessantemente,
ha intenzione di formulare una teoria generale delle lacune.
Infatti, in apertura al manoscritto sono espresse in maniera
esplicita e molto chiara le assunzioni generali a cui Majorana
intende far riferimento, applicando il procedimento della
``quantizzazione dell'equazione di Schr\"odinger'' (ossia quello
di seconda quantizzazione). Come Heisenberg, anche Majorana
ottiene l'espressione della hamiltoniana del sistema in termini
delle lacune, ma \`e evidente l'incompletezza (ovvero
l'interruzione) della teoria nel detto manoscritto.

\`E anche particolarmente interessante la ``giustificazione'' che
d\`a Majorana dell'uso degli anticommutatori piuttosto che dei
commutatori, che viene riferito alla forma particolare della
hamiltoniana e alle corrispondenti equazioni del moto che debbono
essere soddisfatte.

Negli {\it excerpta} Senatore IIb e IIc \footnote{Sebbene si
tratti dello stesso argomento, \`e preferibile introdurre la
divisione in due parti, anzich\`e una sola, in quanto \`e evidente
la non consequenzialit\`a dei due differenti testi.} viene invece
affrontato un tema su cui Majorana torna pi\`u volte nei suoi {\it
Quaderni} di ricerca personali \cite{Quaderni}: se, come ricordato
sopra, la teoria di Maxwell dell'elettromagnetismo deve essere
vista come la meccanica ondulatoria del fotone, allora deve essere
possibile scrivere le equazioni di Maxwell come delle equazioni
del tipo di Dirac per una opportuna funzione d'onda. Un modello
alternativo alla teoria dell'elettrodinamica, basato su una
analogia con la teoria di Dirac dell'elettrone, fu gi\`a avanzato
da Oppenheimer nel 1931 \cite{Opp31}, e sul contributo apportato
da Majorana si \`e gi\`a discusso in altri luoghi
\cite{RecEspGian}, a cui si rimanda per possibili riferimenti
approfonditi sulla seconda parte del manoscritto considerato.

Per gli {\it excerpta} Senatore IIc \`e invece particolarmente
interessante segnalare la presentazione chiara ed esplicita delle
propriet\`a sperimentali dei fotoni su cui si pu\`o basare la
nuova formulazione dell'elettrodinamica, relative alla velocit\`a,
energia, quantit\`a di moto e spin dei fotoni. Si osservi, anche,
che l'equazione del tipo di Dirac per i fotoni riportata alla fine
del IIc \`e sostanzialmente {\it diversa} da quella considerata
nel IIb (e da Oppenheimer), essendo essa applicata ad una funzione
d'onda a {\it due} componenti (corrispondenti agli unici due
diversi stati di polarizzazione del fotone) anzich\`e tre
componenti (corrispondenti ai campi elettrico e magnetico
dell'onda associata).

\section{La composizione del manoscritto}

\noindent Sfortunatamente, gli {\it excerpta} Senatore non sono
datati, per cui risulta molto difficile un'analisi accurata della
genesi e composizione del lavoro di Majorana qui considerato senza
opportune ipotesi. Tuttavia alcune ``ragionevoli'' considerazioni
possono utilmente essere svolte, senza dimenticare il punto
fondamentale appena evidenziato.

Innanzitutto, in base al tipo di scrittura adoperata, \`e
certamente da escludere una data di composizione vicino a quelle
del corso di Majorana a Napoli (1938). Dal confronto con altri
documenti presenti nell'archivio della Domus Galilaeana,
specialmente con gli originali manoscritti degli articoli
pubblicati di Majorana, si ricava infatti che sia il tipo di
scrittura che il tratto della penna utilizzata sono molto simili a
quelli per la stesura dell'articolo {\it Atomi orientati in campo
magnetico variabile} \cite{atomi} del 1932\footnote{Essi sono
anche simili a quelli del manoscritto dell'articolo {\it Sullo
sdoppiamento dei termini Roentgen ottici a causa dell'elettrone
rotante e sulla intensit\`a delle righe del Cesio}
\cite{sdoppiamento} del 1928, indicando quindi una composizione
precedente all'articolo di Heisenberg del 1931 \cite{Hei31} e una
indipendenza da esso. Sebbene tale avanzata ipotesi non possa
completamente escludersi, per quanto si \`e osservato nel
paragrafo precedente e si dir\`a in appresso la si riterr\`a poco
verosimile.}.

L'analisi dei contenuti specifici del manoscritto di Majorana
confermerebbe, poi, una data di composizione non antecedente al
1932 rivelando, per gli {\it excerpta} Senatore IIa, una certa
dipendenza dal'articolo di Heisenberg in \cite{Hei31} e, per gli
{\it excerpta} Senatore IIb e IIc, almeno una conoscenza del
lavoro di Oppenheimer dello stesso anno \cite{Opp31}.

Questi unici dati, tuttavia, portano a delle conclusioni piuttosto
generiche, indicando solo un limite inferiore sulla data di
composizione, e certamente non offrono spunti di riflessione sulla
genesi e il previsto utilizzo del manoscritto.

Ricercando tra tutte le carte lasciate dal fisico catanese, e
conservate sia in archivi pubblici come la Domus Galilaeana che in
quelli privati della famiglia Majorana, si \`e appurato che esiste
un {\it unico} altro documento stilato in francese. Esso
corrisponde alla copia minuta di una lettera di Majorana in
risposta ad un invito ad una conferenza (si veda il documento
MX/R1 in \cite{Recami}), che di seguito si riporta in traduzione:
\begin{quote}
Caro Signore,

vi ringrazio vivamente per il vostro invito di partecipare al
prossimo congresso che avr\`a luogo a Leningrado. Sono felice di
accettare e di avere l'occasione di conoscere al tempo stesso il
vostro grande e bel paese. Ho parlato anche del vostro invito al
Sig. Fermi e al Sig. Rossi. Fermi \`e impegnato per un corso di
conferenze in America e non potr\`a venire. Rossi al contrario mi
ha assicurato che accetter\`a molto volentieri di recarsi in
Russia.

Con i miei vivi ringraziamenti riceviate, caro Signore, i miei
calorosi saluti.

Vostro

{}\hfill{Ettore Majorana}
\end{quote}
Anche tale documento non reca la data di stesura (si osservi che
l'originale della lettera, presumibilmente contenente la data, non
poteva essere pi\`u in possesso di Majorana, ma piuttosto
dell'ignoto destinatario), ma dalle notizie in esso contenute e
dalle indagini effettuate su queste si possono trarre delle
informazioni sufficientemente interessanti.

Innanzitutto, la lettera di invito alla conferenza era rivolta sia
a Majorana che a Bruno Rossi (oltre che a Enrico Fermi), che si
trovava presso l'Istituto di Fisica di Padova (si veda
\cite{Recami}). Ora \`e noto che Rossi era professore a Padova
dall'autunno del 1932, e che a cavallo tra il 1932 e il 1933
collabor\`o direttamente con Fermi (a Roma) su alcune questioni
riguardanti i raggi cosmici (si veda l'articolo in
\cite{RossiFermi} e la presentazione che ne viene fatta in
\cite{FNM} a pag. 509).

Il secondo dato importante \`e quello relativo alla conferenza in
oggetto, che sembra doversi svolgere a Leningrado. Negli anni in
questione, l'unica conferenza che qui interessa \`e la Prima
Conferenza Nucleare dell'Intera Unione \cite{allunion} (il cui
argomento principe fu la fisica del nucleo atomico) che si svolse
dal 24 al 30 settembre del 1933 all'Istituto Fisico-Tecnico di
Leningrado \cite{Vizgin}. Contrariamente agli auspici contenuti
nella lettera di Majorana, a tale conferenza (e ad altre tenutesi
in terra sovietica) non parteciparono n\`e Majorana n\`e Rossi
mentre, tra gli altri, (I.E. Tamm, V.A. Fock, G.A. Gamov, P.A.M.
Dirac, F. Joliot, V.F. Weisskopf) prese parte Franco Rasetti
\cite{Vizgin}, un altro dei ``ragazzi di via Panisperna''. \`E
anche interessante che a tale conferenza partecip\`o Dirac che
tenne un intervento sulla {\it Teoria del positrone}
\cite{positrona}, ossia appunto sulla teoria delle lacune
considerata da Majorana nel suo manoscritto.

La data di svolgimento di tale conferenza andrebbe d'accordo anche
con l'altro importante informazione contenuta nella lettera di
Majorana, riguardo l'assenza di Fermi, impegnato in un ``corso di
conferenze in America''. Infatti, nell'agosto del 1933 Fermi si
rec\`o alla scuola estiva di Ann Arbor \cite{Segre} \cite{FNM77b}
(dove fu invitato a tenere delle lezioni, come anche nel 1930 e
1935) accompagnato da Emilio Segr\'e.

Tuttavia, l'ipotesi or ora considerata non \`e scevra da
difficolt\`a, riguardanti soprattutto: {\it chi} effettivamente
sugger\`i di invitare Majorana e Rossi alla conferenza e {\it
quando} l'invito giunse a Roma e la lettera di Majorana fu
materialmente scritta. Infatti, riguardo al primo punto, sia
l'opera di Majorana che quella di Rossi non erano ancora
sufficientemente conosciute dalla comunit\`a internazionale
all'inizio del 1933, mentre, per il secondo punto, occorre
ricordare che Majorana era a Lipsia (e Copenhagen, ma non a Roma)
dal gennaio fino agli inizi di agosto del 1933 \cite{Recami},
tranne che per le vacanze pasquali di quell'anno (fine aprile -
inizio maggio), per cui difficilmente si poteva organizzare una
visita a Leningrado solo circa un mese prima della conferenza o
nel breve periodo delle vacanze pasquali.

La notoriet\`a internazionale di Majorana nel campo della fisica
nucleare e, in generale, della fisica teorica venne solo dopo il
suo soggiorno a Lipsia da Heisenberg che, tra gli altri, gli fece
molta ``pubblicit\`a'' nelle successive conferenze
\cite{degneutro}, e specialmente nel congresso Solvay di Bruxelles
dell'ottobre 1933, in cui gli interventi erano tenuti in lingua
francese. L'invito a Majorana e a Rossi (lo stesso Heisenberg, che
pure lavor\`o nel campo della fisica dei raggi cosmici, divenne
ugualmente un buon estimatore dei lavori di Rossi) potrebbe quindi
essere stato concepito in questa occasione, e formalizzato
successivamente.

In tale ipotetica evenienza, o anche indipendentemente da essa, la
conferenza a cui ci si riferisce nella lettera non potrebbe essere
quella del settembre 1933 a Leningrado. La successiva conferenza
in suolo russo non si tenne in tale citt\`a, ma nell'altro
importante centro di Kharkov nel maggio del 1934 \cite{Tamm}
\cite{cold}, sebbene organizzata anche da fisici russi di rilievo
che lavoravano a Leningrado, come A.F. Joffe e altri. A tale
Conferenza Internazionale di Fisica Teorica (non specificamente di
fisica nucleare) presero parte pochi personaggi non russi, tra cui
Niels Bohr, Leon Rosenfeld (entrambi conosciuti nel 1933 da
Majorana), Ivar Waller e Walter Gordon, ma \`e interessante notare
che, nell'anno precedente la conferenza, nell'Istituto di Kharkov
lavor\`o il fisico Victor F. Weisskopf \cite{Weisskopf} con Lev D.
Landau e altri, mentre Rudolph Peierls abitualmente faceva visita
a Landau a Leningrado \cite{Peierls}. Sia Weisskopf che Peierls
conoscevano bene Majorana \cite{Recami}, il primo avendolo
incontrato a Lipsia nel 1933 (ed avendo discusso con lui di
elettrodinamica quantistica), mentre il secondo a Roma, verso la
fine del 1932 o l'inizio del 1933, prima che Majorana partisse per
l'estero.

Anche la notizia dell'assenza di Fermi, menzionata nella lettera
di Majorana, pu\`o trovare un'agile collocazione nell'ipotesi di
partecipazione alla conferenza a Kharkov. Infatti, \`e noto che
nell'estate del 1934 Fermi and\`o in Sud America, dove tenne una
serie di conferenze a Buenos Aires, Montevideo e altrove
\cite{Segre}. A voler interpretare letteralmente le parole di
Majorana, il ``corso di conferenze'' sembra pi\`u riferirsi al
1934 in Sud America che non al 1933 negli Stati Uniti.

In definitiva, le ipotesi verosimili sulla occasione di scrittura
della citata lettera di Majorana sono solo due: o la conferenza di
fisica nucleare a Leningrado del settembre 1933, o quella di
fisica teorica a Kharkov del maggio 1934.

La questione che ora si pone \`e quale relazione potrebbe
sussistere tra la lettera sull'invito ad una conferenza in Russia
e la stesura degli {\it excerpta} Senatore II. Purtroppo, a tale
riguardo, nessun dato probante esiste. Tuttavia, non risultando
nella biografia di Majorana \cite{Recami} alcun altra occasione
che possa implicare l'uso della lingua francese come tramite di
comunicazione, sembra ragionevole ammettere un legame stretto tra
i due documenti. Se tale congettura fosse verificata, allora si
potrebbe certamente concludere che il testo presente negli {\it
excerpta} Senatore II sia servito in preparazione all'intervento
di Majorana alla conferenza di Leningrado o a quella di Kharkov,
alla quale, tuttavia, egli non partecip\`o per un ignoto motivo (e
la mancata partecipazione potrebbe anche mettersi in relazione al
non compiuto sviluppo della teoria presente negli {\it excerpta}
che, come accennato sopra, risulta solo abbozzata).

\section{Conclusioni}

\noindent Da una recente ricognizione alla Domus Galilaeana a Pisa
si \`e accertato che nel fascicolo consegnato nel marzo 1938 da
Ettore Majorana ad una sua studentessa a Napoli prima di
scomparire, e contenente principalmente gli appunti delle lezioni
del corso di fisica teorica tenuto dal fisico catanese,
effettivamente sono presenti alcuni fogli non facente parte della
collezione dei suddetti appunti. La notizia dell'esistenza di tali
estratti fu riportata direttamente dalla studentessa di Majorana,
Gilda Senatore, alcuni anni or sono, ma finora non era mai stata
accertata la loro reale presenza nel fascicolo. Nel presente
lavoro \`e stato, quindi, compiuta per la prima volta un'analisi
accurata relativa al manoscritto suddetto ({\it excerpta} Senatore
II) composto da Majorana in lingua {\it francese}. Esso \`e
l'unico testo scientifico da lui redatto in tale lingua. Il
contenuto del detto documento elabora su un abbozzo di teoria
riguardante l'elettrodinamica quantistica, con l'uso del
formalismo della quantizzazione dei campi, e in particolare
sviluppa su alcune questioni riguardanti la teoria delle lacune e
la formulazione dell'elettrodinamica suggerita da Oppenheimer in
analogia alla teoria relativistica di Dirac dell'elettrone.

Da quanto discusso in questa sede, sembra ragionevole supporre che
il testo considerato qui sia stato elaborato nel 1933-1934,
probabilmente in vista di una conferenza in Unione Sovietica (a
Leningrado nel 1933 o a Kharkov nel 1934) a cui Majorana fu
invitato, anche se, effettivamente, non vi partecip\`o. Inogni
caso, si pu\`o plausibilmente escludere una composizione molto
distante dal periodo in cui egli soggiorn\`o a Lipsia da
Heisenberg, essendo ben documentata una dipendenza dello scritto
di Majorana da un articolo dello stesso fisico tedesco.

D'altra parte, come attesta anche Weisskopf, che conobbe Majorana
a Lipsia, $\ll$a quel tempo io ero ancora interessato
all'elettrodinamica quantistica, e c'erano due problemi da
affrontare: uno era il problema del positrone, se veramente esso
era contenuto nell'equazione di Dirac (come diremmo oggi, il
problema della simmetria per coniugazione di carica), e l'altro
era il problema della forza nucleare, l'inizio della fisica
nucleare... Tutte le discussioni si focalizzavano sulla struttura
nucleare da un lato e sull'elettrodinamica quantistica
dall'altro$\gg$ \cite{Weisskopf}.

Majorana, quindi, come gli altri brillanti fisici teorici del
periodo, partecip\`o attivamente ad entrambe le questioni
fondamentali \cite{kern} \cite{elpos}, e gli {\it excerpta}
Senatore II ne sono una ulteriore interessante testimonianza.
Rimane tuttavia il fatto che la teoria sviluppata in tali estratti
non \`e completa, e al momento non \`e possibile congetturare sul
perch\`e il grande fisico teorico non abbia portato a compimento
il suo lavoro. Certamente \`e intrigante che Majorana aveva cone
s\`e a Napoli, ancora a distanza di anni, questi fogli, e che li
consegn\`o alla Senatore, insieme agli appunti delle lezioni del
suo corso, prima di scomparire. Probabilmente, ulteriori indagini
in tale direzione potranno in futuro gettare nuova luce su questo
nuovo e interessante problema.

\section*{Ringraziamenti}

\noindent Il costante interesse e il proficuo aiuto fornitomi
%che mi \`e stato dato, durante lo svolgimento del presente lavoro,
da E. Recami e A. De Gregorio meritano la mia sincera gratitudine,
che viene qui pienamente espressa.

%\newpage

\appendix

\section{Il testo di Majorana}

\noindent Nel seguito viene riportato, tradotto in lingua
italiana, il testo di Majorana presente negli {\it excerpta}
Senatore II. Per evidenti motivi, si \`e preferito eseguire una
traduzione letterale dal francese. Tuttavia va segnalato che
l'impressione che se ne ricava \`e che l'autore abbia forse prima
scritto il testo in italiano e poi lo abbia tradotto in francese,
o comunque non abbia elaborato il testo direttamente seguendo le
regole di composizione in lingua francese. In pochi luoghi, ove
strettamente necessario, eventuale testo non presente nel
manoscritto originale \`e stato aggiunto tra parentesi [...] per
facilitare la comprensione del documento.

\bigskip \bigskip

\centerline{\sc {\it Excerpta} Senatore IIa}

\bigskip

\noindent Consideriamo un sistema di $p$ elettroni e poniamo le
seguenti ipotesi: 1) che l'interazione fra le particelle sia
abbastanza piccola per poter parlare di stati quantici
individuali, considerando che i numeri quantici che definiscono la
configurazione del sistema siano dei buoni numeri quantici; 2) che
ogni elettrone possegga un numero $n>p$ di livelli [energetici]
profondi, mentre tutti gli altri livelli siano molto pi\`u
elevati. Deduciamo che gli stati del sistema tutto intero possono
essere divisi in due classi, la prima costituita da tutte le
configurazioni nelle quali tutti gli elettroni sono in uno degli
stati profondi, mentre appartengono alla seconda tutte le
configurazioni nelle quali almeno \underline{un} elettrone si
trovi in uno stato elevato al di fuori degli $\underline{n}$ gi\`a
menzionati. Supporremo inoltre che sia possibile con sufficiente
approssimazione trascurare le interazioni tra gli stati delle due
classi, cio\`e trascureremo gli elementi di matrice dell'energia
che corrispondono all'accoppiamento di stati di classi differenti,
di modo che ci sar\`a possibile di considerare il movimento delle
$p$ particelle negli $n$ stati profondi, come se questi solo
esistessero. Ebbene, ci proponiamo di ricondurre questo problema a
quello del movimento di $n-p$ particelle negli stessi stati,
queste nuove particelle rappresentando le lacune,\footnote{La
traduzione letterale dal francese darebbe ``posti vuoti''.}
secondo il principio di Pauli.

Per raggiungere il nostro scopo conviene servirci del formalismo
di cui si fa uso nel procedimento detto della quantizzazione
dell'equazione di Schr\"odinger. Consideriamo l'equazione di
Schr\"odinger:
\[
\dot{\psi} = - \frac{2 \pi i}{h} \, H \psi
\]
come definente un campo classico, tenendo conto che intendiamo che
$\psi$ rappresenti non una sola particella, ma al contrario un
numero molto grande di particelle tale da poter trascurare la
costituzione granulare della materia. Allora nel potenziale che
compare nell'hamiltoniano \`e bene porre in evidenza questa parte
che dipende dall'interazione mutua degli elettroni.
%fine pagina
Poniamo dunque: %
\be \label{1} %
- \frac{h}{2 \pi i} \dot{\psi} = H \psi + V \psi
\ee %
essendo \setcounter{equation}{0}
\renewcommand{\theequation}{\arabic{equation}$'$}
\be \label{1p}%
V(P) = \int G(P,P') \,\, \psi^\ast(P') \psi(P') \, \drm \tau
\ee %
\setcounter{equation}{1}\renewcommand{\theequation}{\arabic{equation}}dove
$G(P',P)$ \`e il potenziale delle forze che agiranno tra due
particelle situate in $P$ e $P'$. In modo naturale riguarderemo
come energia del campo l'espressione %
\bea %
& & \int \psi^\ast \, H \, \psi \, \drm \tau + \frac{1}{2} \int
\psi^\ast \, V \, \psi \, \drm \tau \nonumber \\
& & = \int \psi^\ast \, H \, \psi \, \drm \tau + \frac{1}{2} \int
\!\!\!\! \int G(P,P') \, \psi^\ast(P) \, \psi^\ast(P') \, \psi(P)
\, \psi(P') \, \drm \tau \, \drm \tau' \label{2}
\eea %
Sviluppiamo ora secondo un sistema di funzioni
ortogonali:\footnote{Le equazioni che compaiono nel secondo e
terzo rigo seguenti non sono scritte nel manoscritto di seguito
alla prima equazione, ma nell'angolo libero in alto nel foglio.
Nella seconda equazione \`e utilizzato il simbolo $\psi_k$ invece
che $\varphi_k$.}
\[
\psi = \sum a_i \, \varphi_i \qquad \qquad \qquad \int
\varphi_i^\ast \, \varphi_k \, \drm \tau = \delta_{ik}
\]
\[
H \varphi_k = \sum_i H_{ik} \, \varphi_i
\]
\[
V_{ik} = \frac{1}{2} \sum_{\ell , m} O_{i \ell, k m} \,
a^\ast_\ell a_m
\]
possiamo scrivere:\footnote{Nel manoscritto l'equazione
seguente \`e scritta come \[ H_{ik} = \int \widetilde{\psi}_i \, H
\, \psi_k \, \drm \tau\]}
\[
H_{ik} = \int \varphi^\ast_i \, H \, \varphi_k \, \drm \tau
\]
\[
O_{i \ell , k m} = \int \!\!\!\! \int G(P,P') \, \varphi^\ast_i(P)
\, \varphi^\ast_\ell(P') \, \varphi_k(P) \, \varphi_m(P') \, \drm
\tau \, \drm \tau'
\]
Allora sostituendo nelle equazioni del moto scriveremo: %
\be \label{3} %
\dot{a}_i = - \frac{2 \pi i}{h} \left\{ \sum_k H_{ik} \, a_k +
\sum_{\ell, k, m} O_{i \ell, k m} \, a^\ast_\ell a_k a_m \right\}
\ee %
e come espressione dell'energia: %
\be \label{4} %
W = \sum_{i, k} H_{ik} \, a^\ast_i a_k + \frac{1}{2} \sum_{i,
\ell, k, m} O_{i \ell, k m} \, a^\ast_i a^\ast_\ell a_k a_m
\ee %
Tenendo presente queste equazioni, diamo ora loro un significato
quantistico, ponendo %
\be \label{5} %
\dot{a}_i = - \frac{2 \pi i}{h} \left( a W - W a \right) \qquad
\qquad \dot{a}^\ast_i = - \frac{2 \pi i}{h} \left( a^\ast W - W
a^\ast \right)
\ee %
le $a$ essendo ora delle matrici. Si vede facilmente che,
affinch\`e le equazioni (\ref{5}) siano equivalenti alle equazioni
(\ref{3}) occorre che le $a$ soddisfino alle relazioni di scambio:
\begin{eqnarray*}
& & a_i a^\ast_k - a^\ast_k a_i = \delta_{i k} \\
& & a_i a_k - a_k a_i = 0 \\
& & a^\ast_i a^\ast_k - a^\ast_k a^\ast_i = 0
\end{eqnarray*}
Ci\`o significa quantizzare secondo le regole classiche di
Heisenberg poich\`e, in effetti, i momenti coniugati alle
variabili $a$ sono classicamente le $a^\ast$ moltiplicate per $-h
/ 2 \pi i$. Le relazioni di scambio di Heisenberg ci porterebbero
a considerare delle particelle che obbediscono alla statistica di
Bose, mentre ci\`o che ci interessa \`e l'altro caso, quello delle
particelle che obbediscono alla statistica di Fermi. Come Jordan e
Wigner hanno dimostrato, occorre per questo cambiare i segni delle
relazioni di Heisenberg:
\begin{eqnarray}
& & a_i a^\ast_k + a^\ast_k a_i = \delta_{i k} \nonumber \\
& & a_i a_k + a_k a_i = 0 \label{6} \\
& & a^\ast_i a^\ast_k + a^\ast_k a^\ast_i = 0 \nonumber
\end{eqnarray}
Ci\`o non \`e giustificabile da un punto di vista generale, ma
solamente dalla forma particolare dell'hamiltoniano; si pu\`o
infatti verificare che le equazioni del moto sono soddisfatte al
meglio da queste ultime relazioni di scambio che non quelle di
Heisenberg. Troveremo ora una soluzione di comodo.%
\[
I_k = \left| \ba{cc} 1 & 0 \\ 0 & 1 \ea \right| \qquad  I'_k =
\left| \ba{cc} 1 & 0 \\ 0 & -1 \ea \right| \qquad \alpha_k =
\left| \ba{cc} 0 & 1 \\ 0 & 0 \ea \right| \qquad \alpha^\ast_k =
\left| \ba{cc} 0 & 0 \\ 1 & 0 \ea \right|
\]

\[
a_i = I'_1 \times I'_2 \dots I'_{i-1} \times \alpha_i \times
I_{i+1} \times I_{i+2} \dots
\]

\[
a^\ast_i = I'_1 \times I'_2 \dots I'_{i-1} \times \alpha^\ast_i
\times I_{i+1} \times I_{i+2} \dots
\]

\[
\alpha^\ast_i \alpha_i = \left| \ba{cc} 0 & 0 \\ 0 & 1 \ea \right|
\]
[i cui autovalori sono] 0,1

\[
\int \psi^\ast \psi \, \drm \tau = \sum_i \alpha^\ast_i \alpha_i =
n
\]

\noindent \footnote{Nel seguito del manoscritto sono esperiti
alcuni tentativi di calcolo degli elementi di matrice dell'energia
$W$ (e, in particolare, di $H_{ik}$ e $O_{i \ell, k m}$) nella
base dei numeri di occupazione $n_1, n_2, \dots, n_i, \dots, n_k,
\dots$ che, tuttavia, qui non riportiamo.}

\noindent [Ponendo] $b_i=a_i^\ast$, $b_i^\ast=a_i$ [la relazione]

\[
a_i^\ast a_i + a_i a_i^\ast = 1
\]

\noindent [pu\`o scriversi]

\[
a_i^\ast a_i + b_i^\ast b_i = 1
\]

\noindent [ovvero]

\[
a_i^\ast a_i = 1 \qquad \qquad b_i^\ast b_i = 0 \hphantom{.}
\]

\noindent [oppure]

\[
a_i^\ast a_i = 0 \qquad \qquad b_i^\ast b_i = 1 .
\]

\noindent [Per l'energia $W$ si ha quindi:]

\[
\sum H_{ik} \, a^\ast_i a_k + \frac{1}{2} \sum O_{i \ell, k m} \,
a^\ast_i a^\ast_\ell a_k a_m
\]

\noindent [oppure]

\[
\sum H_{ik} \, b_i b^\ast_k + \frac{1}{2} \sum O_{i \ell, k m} \,
b_i b_\ell b^\ast_k b^\ast_m .
\]

\noindent [Dalla relazione]

\begin{eqnarray*}
b_i^\ast b_k + b_k^\ast b_i = \delta_{ik} \\
b_i^\ast b_k = \delta_{ik} - b_k^\ast b_i
\end{eqnarray*}

\noindent [il primo termine dell'energia si scrive:]

\[
\sum H_{ii} - \sum \overline{H}_{ik} b_i^\ast b_k .
\]

\noindent [Per il secondo termine, utilizzando:]

\begin{eqnarray*}
b_i b_\ell b_k^\ast b_m^\ast &=& b_i \left( \delta{\ell k}-
b_k^\ast b_\ell \right) b_m^\ast \\
&=& \delta_{\ell k} b_i b_m^\ast - b_i b_k^\ast b_\ell b_m^\ast = \dots \\
&=& b_k^\ast b_m^\ast b_i b_\ell + \delta_{\ell k} b_i b_m^\ast +
\delta_{i m} b_\ell b^\ast_k - \delta_{i k} b_\ell b_m^\ast -
\delta_{\ell m} b_i b_k^\ast
\end{eqnarray*}

\noindent \footnote{Nell'ultima relazione l'autore ha omesso il
termine $\delta_{\ell m} \delta_{ik} - \delta_{i m} \delta_{\ell
k}$, che nella sommatoria che appare nell'espressione dell'energia
porta un contributo nullo.}

\noindent [si ottiene:]

\begin{eqnarray*}
\frac{1}{2} \sum O_{i \ell,km} \, b_i b_\ell b_k^\ast b_m^\ast &=&
\frac{1}{2} \sum O_{i \ell,km} \, b_k^\ast b_m^\ast b_i b_\ell \\
&+& \frac{1}{2} \sum O_{i \ell,\ell m} \, b_i b_m^\ast +
\frac{1}{2} \sum O_{i \ell,k i} \, b_\ell b_k^\ast + \dots
\end{eqnarray*}

\noindent \footnote{Il testo contenuto nel manoscritto termina con
altri pochi calcoli che qui non riportiamo.}

\bigskip \bigskip

\centerline{\sc {\it Excerpta} Senatore IIb}

\bigskip

\noindent Queste\footnote{L'inizio di questa parte del manoscritto
\`e evidentemente il continuo di una parte precedente, andata
persa. Si veda, tuttavia, quanto riportato dall'autore nel
seguito.} equazioni ne soddisfano tuttavia una quarta. \\
Allora Opp.\footnote{Majorana si riferisce qui, probabilmente,
all'articolo di R.J. Oppenheimer in {\it Phys. Rev.} {\bf 38}
(1931) 725.} ha tentato di considerare una teoria a tre componenti
vettoriali.

\noindent [Le equazioni di Maxwell nel vuoto]

\[
\frac{1}{c} \frac{\partial E}{\partial t} = \rot H \qquad
\frac{1}{c} \frac{\partial H}{\partial t} = - \rot E
\]
\[
\div E = 0 \qquad \div H = 0
\]

${}$

\noindent [ponendo]

\begin{eqnarray*}
\psi_x &=& E_x - i H_x \\
\psi_y &=& E_y - i H_y \\
\psi_z &=& E_z - i H_z
\end{eqnarray*}

${}$

\noindent [possono scriversi nella seguente forma:]

\[
\frac{1}{c} \frac{\partial \psi}{\partial t} = \rot \left( H + i E
\right) = i \, \rot \left( E - i H \right) = i \, \rot \psi
\]

\[
\left\{ \ba{l} \displaystyle \frac{1}{c} \frac{\partial
\psi}{\partial t} - i \, \rot \psi = 0 \\ \\ \displaystyle \div
\psi = 0 \ea \right.
\]

${}$

\[
\left\{ \ba{l} \displaystyle \frac{1}{c} \frac{\partial
\psi_x}{\partial t} - i \, \frac{\partial \psi_z}{\partial y} + i
\, \frac{\partial \psi_y}{\partial z} = 0
\\ \\
\dots \\ \\
\displaystyle \frac{\partial \psi_x}{\partial x} + \frac{\partial
\psi_y}{\partial y} + \frac{\partial \psi_z}{\partial z} = 0 \ea
\right.
\]

${}$

\noindent [Introducendo gli operatori]

\[ W = - \frac{h}{2 \pi i} \, \frac{\partial ~}{\partial t}, \qquad
p_x = \frac{h}{2 \pi i} \, \frac{\partial ~}{\partial x}, \qquad
\dots \]

${}$

\noindent [si ha:]

\[
\left\{ \ba{l} \displaystyle \frac{1}{c} W \psi_x + i \, p_y
\psi_z - i \, p_z \psi_y = 0
\\ \\
\displaystyle \frac{1}{c} W \psi_y + i \, p_z \psi_x - i \, p_x
\psi_z = 0
\\ \\
\displaystyle \frac{1}{c} W \psi_z + i \, p_x \psi_y - i \, p_y
\psi_x = 0 \ea \right. \qquad \qquad \underline{p_x \psi_x + p_y
\psi_y + p_z \psi_z = 0}
\]

${}$

\noindent [e le equazioni di Maxwell possono scriversi nella forma
compatta, analoga a quella dell'equazione di Dirac,]

\[
\left[ \frac{1}{c} \, W + \left( \alpha , p \right) \right] \psi =
0
\]

${}$

\noindent [con]

\[ \alpha_x \; = \; \left( \ba{ccc}
0 & 0 & 0 \\
0 & 0 & -i \\
0 & i & 0 \ea \right) \qquad \alpha_y \; = \; \left( \ba{ccc}
0 & 0 & i \\
0 & 0 & 0 \\
-i & 0 & 0 \ea \right) \qquad \alpha_z \; = \; \left( \ba{ccc}
0 & -i & 0 \\
i & 0 & 0 \\
0 & 0 & 0 \ea \right)
\]

${}$

\noindent [Affinch\`e l'equazione a cui soddisfa $\psi$ abbia
soluzione non banale, occorre che l'energia $W$ sia data da]

\[
\frac{W}{c} = \left( \ba{ccc}
0 & i p_z & -i p_y \\
-i p_z & 0 & i p_x \\
i p_y & -i p_x & 0 \ea \right)
\]

${}$

\noindent [i cui autovalori sono]

\[
\frac{W}{c} =  p \qquad \frac{W}{c} = - p \qquad \frac{W}{c} = 0
\]

${}$

\noindent [mentre il rapporto tra le componenti dell'autovettore
corrispondente all'autovalore nullo \`e dato da]

\[
\psi_x : \psi_y : \psi_z \; = \; - p_x^2 : -p_x p_y : -p_x p_z \;
= \; p_x : p_y : p_z
\]

\[
\widetilde{\psi} \psi = E^2 + B^2
\]

\bigskip \bigskip

\centerline{\sc {\it Excerpta} Senatore IIc}

\bigskip

\noindent Oppenheimer ha cercato di costruire queste equazioni con
un procedimento induttivo servendosi delle propriet\`a
sperimentali dei fotoni. \\
Queste propriet\`a possono essere riassunte nella seguente
maniera: \\
1) I fotoni si muovono con la velocit\`a della luce \\
2) L'energia e la quantit\`a di moto sono legate dalla relazione
molto semplice: $W = c p$ \\
3) In base alla grandezza e direzione della quantit\`a di moto
sono ancora possibili due stati diversi di polarizzazione \\
4) Il fotone possiede un momento intrinseco che pu\`o assumere i
valori $\displaystyle \pm \frac{h}{2 \pi}$ attorno all'asse di
propagazione. Per la giustificazione sperimentale di quest'ultimo
postulato, osserviamo che, quando un atomo emette della
radiazione, il suo momento angolare attorno a un asse $z$ cambia
da $+$ a $\displaystyle - \frac{h}{2 \pi}$, oppure di $0$, secondo
che ha luogo l'una o l'altra delle transizioni:
\[
m \rightarrow m^\prime = m \mp 1
\]
\[
\left( m \rightarrow m^\prime = m \right)
\]
Ebbene, se si osserva un fotone lungo l'asse $z$, si pu\`o
certamente dire che questo fotone non \`e dovuto all'ultima
transizione, perch\`e l'intensit\`a di quest'ultima transizione si
annulla lungo quest'asse. D'altra parte se il fotone si trova
esattamente sull'asse $z$, ci\`o non pu\`o dipendere dal momento
orbitale attorno a questo asse, consegue che per la conservazione
del momento angolare del sistema, si deve ammettere l'esistenza
del momento intrinseco del fotone, e precisamente tale che sia
capace dei due valori $\displaystyle \pm \frac{h}{2 \pi}$ secondo
la direzione di propagazione. Questa maniera di interpretare lo
spin del fotone \`e stata proposta per la prima volta da Dirac,
quando egli non era ancora R.R.S.\footnote{Probabilmente tale
sigla andrebbe corretta in F.R.S., {\it Fellow of the Royal
Society}, intendendo che il lavoro di Dirac risale a prima che
egli fosse eletto membro della Royal Society di Londra (nel marzo
1930). \`E comunque interessante trovare una siffatta nota storica
in un manoscritto scientifico di Majorana.} \\
Poich\`e ci troviamo di fronte  a due componenti polarizzate,
sarebbe bene avere una teoria ondulatoria a due componenti.
Infatti si possono costruire delle equazioni della forma
desiderata che soddisfano ai tre primi postulati. Queste equazioni
sono le seguenti:
\[
\left[ \frac{W}{c} + \left( \sigma , p \right) \right] \psi = 0
\qquad \qquad \sigma_x = \left| \ba{cc} 0 & 1 \\ 1 & 0 \ea \right|
\quad \sigma_y = \left| \ba{cc} 0 & -i \\ i & 0 \ea \right| \quad
\sigma_z = \left| \ba{cc} 1 & 0 \\ 0 & - 1 \ea \right|
\]
Faccio notare che queste equazioni non sono altro che la met\`a
delle equazioni di Dirac.

%-----------------------------------------------------------------

%-----------------------------------------------------------------

\end{document}